# Intelligent Usability Evaluation for Fashion Websites

**Asmaa Hakami [1], Raneem Alqarni [1], Asmaa Muqaibil[1] and Nahed Alowidi [1,\***

[1] Department of Computer Science, Faculty of Computing and Information Technology, King Abdulaziz University, Jeddah (21589), Saudi Arabia.
\* Correspondence: nalowidi@kau.edu.sa.

**Abstract**

Websites have become increasingly important in people's lives, fulfilling a wide range of needs across various domains such as shopping, education, news, and booking. Among the most heavily used website categories are online shopping platforms, whose usage has particularly increased during the COVID-19 pandemic, as they eliminate time and geographical barriers, providing access to a broader customer base. For these websites to effectively meet user needs and deliver a positive experience, they must be well-designed and adhere to usability principles. However, some existing shopping websites are poorly designed and do not follow usability best practices, resulting in suboptimal user experiences. Traditional manual website evaluation methods are time-consuming, and there is a need for more intelligent, automated approaches, particularly those leveraging machine learning techniques. This study aims to assist fashion shopping website developers in improving the usability of their platforms by providing an intelligent approach that can evaluate website usability. The study employs two complementary approaches for the evaluation process. The first model utilizes a Support Vector Machine (SVM) to assess websites based on specific usability principles, while the second model is a Convolutional Neural Network (CNN) that evaluates websites using features extracted from their screenshot images. The datasets for this project were custom-built, comprising a textual dataset for the SVM model and a screenshot dataset for the CNN model. The results demonstrate that the SVM model achieved an impressive 99% accuracy, while the CNN model attained 69% accuracy. These findings highlight the potential of this intelligent approach to provide comprehensive, data-driven insights for improving the usability of fashion shopping websites.

## 1. Introduction

Internet websites, now-a-days, play an important role in people's life. They help them to meet their needs more easily while saving their time and effort. They have grown rapidly in several fields, such as shopping, education, news, booking, etc. For commerce companies' owners, it is notable that many of them tend to have a website for their companies. Websites are considered as the center of the companies' online presence, and often are the first option for commerce companies to offer their services and products through the Internet for a wide range of customers.

One of the websites that people most use is shopping websites. People prefer using shopping websites as they can buy products from several brands and companies that are not available for purchase in their home countries, they are not bound by a specific time for shopping, and they can purchase items by simply sitting in their homes. Moreover, due to the COVID-19 pandemic, the usage of websites especially shopping web- sites has increased to meet people's needs while maintaining safety. Websites in general and shopping websites specifically must be properly designed, and the developers must pay more attention to this, since the design of the websites gives the first impression of the business, builds trust with the users, and makes them compete with many other websites in the same fields. Well-designed websites provide good usability, and good usability binds more people. According to the new version of International Standards Organization's (ISO 9241-11), the usability term is defined in general as

"the extent to which a system, product or service can be used by specified users to achieve specified goals with effectiveness, efficiency and satisfaction in a specified context of use" [1].

Shopping websites are developed to facilitate meeting users' needs and achieving their goals successfully. Unfortunately, sometimes this aim cannot be achieved due to some existing shopping websites that are designed poorly and do not follow usability principles. This issue might result in a bad user experience, unsatisfied users, and even business loss. For example, if a user visits a shopping website for the first time and finds it difficult to find what he is looking for or gets confused due to the inappropriate distribution of graphical interface elements, he might not visit or use this website again. To have well-designed websites, the usability of the websites should be evaluated, and the evaluation results should be provided to the website developers to help them improve their websites. Unfortunately, websites' usability is often evaluated manually, which are time and effort-consuming processes. Moreover, there is a lack of intelligent evaluation methods, particularly methods that use machine learning.

This work proposes two intelligent models for fashion shopping website usability evaluation. Deep learning (DL), particularly Convolutional Neural Network (CNN) to evaluate the websites by extracting the features from website screenshot and evaluates the website according to them, and machine learning, particularly Support Vector Machine (SVM) to evaluate the websites according to specific usability principles. These models seek to help web developers to identify usability issues and violations that might exist in the design of their website, by showing the locations of the violations and provides some suggestions to the developers to improve the design and the usability of their websites.

The organization of this paper is as follows: Section 2 reviews the relevant prior work in this domain. Section 3 provides a foundational overview of website usability. Section 4 details the dataset employed in the study. Section 5 outlines the performance metrics used to evaluate the models. Section 6 presents a detailed description of the classification models. The paper concludes in Section 7, with potential avenues for future research explored in Section 8.

## 2. Related Work

This section review examines existing research on both non-intelligent and intelligent methods for evaluating website usability, highlighting their strengths, weaknesses, and contributions to the field.

*Non-Intelligent Usability Evaluation Methods*

Traditional usability evaluation methods primarily rely on human experts to assess website quality. Studies such as [2] have employed the Kano model and fuzzy sets theory to classify website features based on customer expectations. By analyzing user feedback through questionnaires, researchers identified key attributes influencing usability. While these methods provide valuable insights, they are inherently subjective and time-consuming. Nizamani et al. [3] adopted a similar approach by developing a set of usability guidelines for university websites. Expert evaluations were conducted to rank websites based on these criteria. Although this method offers a structured approach, it remains reliant on human judgment and may be susceptible to biases.

*Intelligent Usability Evaluation Methods*

In contrast to manual evaluation, intelligent methods leverage machine learning to automate the usability assessment process. Dingli and Cassar [4] pioneered the use of natural language processing techniques to analyze website content, focusing on headings, graphics, and homepage structure. While their approach demonstrated potential, it faced limitations in accurately identifying mixed image types and determining heading levels. Jayanthi and Krishnakumari [5] proposed a different approach using the Extreme Learning Machine (ELM) to classify web pages as good or bad based on features like word count, reading complexity, and link density. While ELM offers advantages in terms of speed and accuracy, its reliance on a limited set of features may restrict its predictive power. Khani et al. [6] explored the use of deep learning for website aesthetic evaluation. By employing Convolutional Neural Networks (CNNs), they aimed to capture visual features that influence user perception. While their model showed promise, its performance was influenced by cultural factors. Lu et al. [7] focused on automated software GUI testing using deep learning. Their approach involved capturing screenshots, training DNN models for classification, and achieving high accuracy in identifying positive and negative GUI elements. Jafary et al. [8] combined qualitative and quantitative methods to evaluate news website quality. While the Delphi technique was used to identify key metrics, machine learning models (MLP and ANFIS) were employed for prediction. Dou et al. [9] used deep convolutional neural networks to assess website aesthetics based on user ratings. Their model demonstrated strong performance in predicting user preferences. Ghattas and Sartawi [10] applied linear regression and support vector machine regression to evaluate website performance based on various metrics. Their findings indicated the superiority of linear regression in terms of accuracy and computational efficiency.

All of the previously mentioned works have utilized AI techniques and methods for automatic website evaluation. They evaluated the websites for different purposes including aesthetics, usability, quality, and performance. However, none of the previous works have evaluated shopping websites specifically, whereas fashion shopping websites are very commonly used and play an important role in the fulfillment of people's needs, in particular during the COVID-19 period. In addition, there is a lack of works regarding usability evaluation of websites. Only one work evaluates websites' usability based on machine learning techniques. This work has not used Deep Learning, which is widely used in solving complex problems and has achieved outstanding accurate results. Moreover, most of the previous works do not provide recommendations for web developers to improve the website.

This paper aims to build two intelligent models using, Convolutional Neural Network (CNN) and Support Vector Machine (SVM) algorithms for usability evaluation of shopping websites. It will evaluate fashion shopping websites according to specific usability principles and provide recommendations for improvement to the developers. The usability principles of the SVM model are load time, mobile- UI, high-resolution photos, and contact information. The CNN model extracts the principles from website screenshots. The usability principles are determined based on the studies [2 and 11] as they consider one of the most significant principles for the target user. There are two datasets, screenshot dataset, and textual dataset.

## 3. Website Usability

Effective website design prioritizes simplicity, ease of use, and consistency to enhance user experience. However, creating a universally applicable design pattern proves challenging due to the diverse nature of websites. Usability, as a design approach, focuses on assessing the ease with which users can learn and interact with a system. It is crucial to recognize that users often possess familiarity with certain graphical user interface (GUI) conventions and deviating from these conventions can hinder usability. For instance, the underlining of hyperlinks is a widely recognized convention that aids in user navigation [12].

### 3.1 Website Usability Principles

Usability principles are a form of guidance and known as heuristics when applied in practice. They evaluate the acceptability of interfaces and often used as the basis for evaluating prototypes and existing systems [13]. Moreover, the criteria provide guidelines to the designers for restricting the space of design alternatives to prevent the designers from designing unusable products [14].
The most commonly used usability principles are developed by Jakob Nielsen [15]. Nielsen's usability principles include visibility of system status, match between system and the real world, user control and freedom, consistency and standards, error prevention, recognition rather than recall, flexibility and efficiency of use, aesthetics and minimalist design, help users recognize, diagnose and recover from errors, help and documentation [15].

**3.2 Usability Principles of E-commerce Website**

Singh and his colleagues [11] have evaluated the usability of ecommerce websites based on six usability parameters including user satisfaction, attractiveness, simplicity, speed, efficiency, searching, and product information. Each of these parameters has several sub-parameters, as shown in Figure 1. User satisfaction is defined as the measure of the quality of a company's service and products. Simplicity in a website is seen in regarding to its operations. Attractiveness involves designing an attractive website that attracts more users by providing high-quality services. Speed is all about managing the time required to load a web page. Efficiency is related to the successful running of a website to meet the users' needs. Searching is about efficient product search result. Finally, product information involves providing the correct details related to the product.

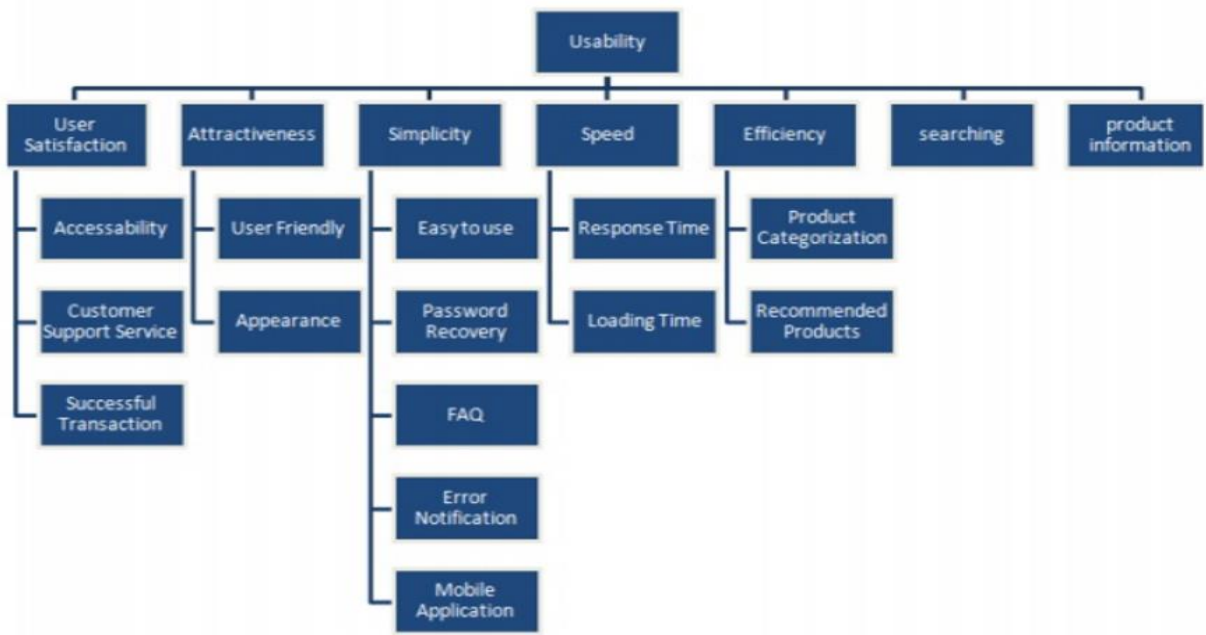

Figure 1: Hierarchical Chart of Usability Parameters and Sub Parameters.
Source: (Singh et al., 2016)[11]

In addition, another study done by Ilbahar and Cebi [2] identified twenty-two parameters for evaluation e-commerce websites. The parameters are: a compare with option, live support, access to customer representative via telephone, customer comments, ability to purchase without signing in, 3D secure, availability of customer comments on the website, the existence of product rates, search in the website option, sort by price option, sort by customer satisfaction score option, sort by sales amount option, high-resolution photo and photo magnification, shopping in a different language, different payment options, information on product features, information on product delivery, access to the page through applications, saving address and contact information for future operations, access to the website through a common account (Google, Facebook, etc.), sort by product features option, and cargo integrated product tracking system.

## 4. Dataset

The dataset for this work was collected and labeled from scratch because there was no available dataset for fashion shopping websites that contain the same principles for evaluation.

**4.1 Textual Dataset**

*Extraction usability principles Challenges*

Nine usability principles have been chosen as principles to evaluate the websites as usable or not. The nine features were load time, mobile-UI, customer comments, the existence of product rate, sorted by many options, high-resolution photo, supported by different languages, different payment options, and contact information. During the implementation phase, these features have been reduced to be four. Only these four features can be extracted automatically either from the HTML source code by a fixed structure or from Web driver. The rest of the features can't be extracted because they do not have a fixed structure and the id of them differ from one website to another. So, they were excluded.

*Extracted usability principles Description*

Load time: One of the most important features that influence customer satisfaction in the usability evaluation of any shopping website is the load time. The load time means the time taken to load the whole page in the browser from a user clicks a link or makes a request. The load time is affected by many factors like web page design. Also, the same website could have different load times depend on the browser, device, geographical location of the user, and even from page to page. Webdriver library has been used to get the URL of the website that needs to measure load time for it. The navigationStart and domComplete time are calculated using window.performance.timing and then take the difference between them. The result of the difference is divided by 1000 to measure the time in seconds.

Mobile-UI: A web page is mobile-UI if the display of its components appropriate for a smartphone or tablet. There is a line in the code indicating the page as a mobile-UI or not, and this line starts with word meta. All lines that start with meta retrieved and their contents or name examined to verify if it has the phrases that refer to the page as mobile-UI or not. Three phrases were used in the code which are "device-width" "apple-mobile-web" and "inmobi-site-verification" if these phrases existed then the page is mobile-UI. These phrases were chosen because when some of the HTML codes were checked these were the most repeated phrase.

High Resolution Photo: The image resolution is mean that how much detail is represented in an image. High resolution means more details which are needed in shopping websites. It also affects load time. The optimization and format of images are used to measure the resolution. The values obtained by using requests, response, and JavaScript Object Notation (JSON), to take the result from https://www.googleapis.com/pagespeedonline/v5/runPagespeed by sending the URL of the website that needs to evaluate optimization and format for its images. The result is normalized values between 0 - 1, the 1 mean the best, and 0 mean the worst. The summation of the two values was taken and divided by 2 to take the average of the resolution. If the result between 1 - 0.8, it will be labeled as A, if it between 0.79 - 0.6, it will be labeled as B, if it between 0.59 - 0.4, it will be labeled as C, if it between 0.39 - 0.2 it will be labeled as D, and the rest will be labeled as F which is the worst.

Contact Information: The contact information in websites is any way to communicate with the website owner or responsible people, such as email, phone, Twitter. The HTML source code is needed to extract contact information. Also, regular expressions should make for each contact ways. So, we go through the HTML source code and find all values that belong to our regular expression.

All these features will be extracted using the web-scraping technique

### 4.2 Screenshot Dataset

To capture comprehensive visual representations of websites for analysis, full-page screenshots were obtained using Selenium WebDriver. This approach ensured that the entire webpage content, including elements that may not be immediately visible within the viewport, was captured for subsequent image processing and analysis.

### 5. Performance Metrics

The performance was comprehensively evaluated using a suite of well-established evaluation metrics [16].

Accuracy represents the overall correct classification rate.

$$Accuracy = \frac{(TP + TN)}{(TP + TN + FP + FN)} \quad (1)$$

Precision measures the proportion of correctly predicted positive instances among all predicted positives.

$$Precision = \frac{TP}{(TP + FP)} \quad (2)$$

Recall indicates the proportion of actual positive instances correctly identified.

$$Recall = \frac{TP}{(TP + FN)} \quad (3)$$

The F1-score, calculated as the harmonic mean of precision and recall, provides a balanced evaluation of the model's performance.

$$F1 - Score = 2\left(\frac{Precision \times Recall}{Precision + Recall}\right) \quad (4)$$

### 6. Classification Models

To assess fashion website usability, two distinct classification models were developed: a textual dataset-based model utilizing Support Vector Machines (SVM) and a screenshot-based model employing Convolutional Neural Networks (CNN). SVM's proficiency in text classification and CNN's strength in image feature extraction justified their selection. The textual model relied on predefined features extracted from websites, while the CNN model learned features directly from screenshot images. Both datasets were meticulously constructed from scratch as no pre-existing fashion website datasets were available.

## 6.1 Textual Dataset-Based Classification Model (SVM Classifier)

*Dataset Collection*

In the textual dataset, 422 shopping websites' URLs were collected. Each website is described by features that represent the usability principles of evaluation. These different features are extracted and determined for each website. The features are load time, mobile-UI, high-resolution photos, and contact information. The value of each feature varies from one feature to another. Two features which are mobile-UI and contact information take categorical data values either yes or no. Where yes indicates the existence of this feature and no indicates the nonexistence of this feature. Load time feature takes numerical value represents the time in seconds. Finally, high-resolution photos take a categorical data value that represents a grade scale from A indicating the most adherence to this feature to F indicating the least adherence to the feature. Grade A for excellent, B for very good, C for good, D for bad, and F for very bad resolution. The features of websites in the dataset are extracted using python code for extracting them automatically.

*Dataset Preprocessing*

Encoding: K-mean cluster algorithm doesn't allow categorical values. So, all the categorical features in the textual dataset are converted to numerical using LabelEncoder.

Normalization: The load time feature has high values. Moreover, the high-resolution photo has five values where the rest two features have binary values. So, all features are normalized using StandardScaler.

*Dataset Labeling*

For each website, there is an attached usability evaluation grade which is the class or the label. The defined five classes of websites are excellent, very good, good, bad, and very bad. These classes are determined for each website by using the clustering technique, which is the K-means algorithm. K-Means build-in function is used with five clusters to fit the dataset. Five clusters were produced ranging from 0 to 4. For deciding the correct descriptive label of each cluster, the importance of the features was taken into consideration. The importance is determined from the results of the questionnaire conducted in the analysis phase. The questionnaire respondents gave several different criteria for the following question "If you had an opportunity to evaluate your website using a smart tool, what criteria would you prefer to use?". The results showed that fifteen respondents said speed, six respondents said high-resolution image, three respondents said responsive layout, and one respondent said communication ways. Based on these results, the ranking of the features from the most important one to the least important is as follows load time, high-resolution image, mobile-UI, and contact information. Clusters 0, 3, 2, 4, and 1 became represents excellent, very good, good, bad, very bad respectively.

*Dataset Splitting*

To make the model able to classify better, it needs to be trained and tested on a labeled data. In this project, the dataset was divided by the train test split method at the rate of 70:30 for training and testing. This method is considered a library in Python language.

*Model Implementation*

The model was programmed by python language, and it was used to classify websites according to their textual data. The parameters of the SVM C and Gamma were selected by the GridSearch method because there is no rule of thumb to choose a C, and Gamma value. This method has been used to choose the best value of C and Gamma among the given values. The given values were C: [0.1, 1, 10, 100, 1000], Gamma: [1, 0.1, 0.01, 0.001, 0.0001], and best values were C=1000, and Gamma=0.001. After choosing the best values, the model was trained and tested on previously collected textual data, and then the accuracy, precision, and recall of it were measured and display.

*Model Performance and Discussion*

The results of the SVM model evaluation are presented in Table 1. The model achieved an impressive overall accuracy of 99%. Additionally, the average Precision, Recall, and F1-Score were all exceptionally high, reaching 100%, 99%, and 99% respectively.

Table 1: SVM Model Performance Metrics

| Accuracy | Precision | Recall | F1-Score |
|---|---|---|---|
| 99% | 100% | 99% | 99% |

These results indicate that the SVM model was able to correctly classify the usability of the fashion websites in the vast majority of cases. The near-perfect Precision score suggests that the model rarely misidentified non-usable websites as usable. Similarly, the high Recall score demonstrates the model's ability to consistently identify truly usable websites. The F1-Score, which balances Precision and Recall, reflects the overall outstanding performance of the SVM model for this task.

The combination of the exceptionally high accuracy, Precision, Recall, and F1-Score suggests that the SVM model was highly effective in classifying the usability of fashion websites based on the provided textual features. These findings demonstrate the suitability of the SVM algorithm for this type of website usability assessment problem, where accurate classification is crucial.

**6.2 Screenshot Dataset-Based Classification Model (CNN Classifier)**

*Dataset Collection*

The websites in the screenshot dataset were intended to be collected for the same websites as the textual dataset, but several websites had problems with their screenshots or with their evaluation. These websites are excluded from the websites' screenshot dataset because the screenshots of them were not good screenshot images. Besides, the evaluation tool used to evaluate the websites could not reach these websites and evaluate them because of denied access or security protection. As a result, several new websites were collected to increase the number of images in the dataset. This dataset consists of 310 screenshots captured from different shopping websites' homepages. These screenshots were captured using either free online tools or built-in command in the Chrome browser for capturing screenshots. These tools are Site-Shot, Screenshot Guru, and Geekflare Screenshot Checker. Tools were used because they can capture a screenshot for lengthy web pages with high quality. In this dataset, the features for each website are not determined taking the advantage of the CNN to extract them by itself from the screenshots images.

*Dataset Prepossessing*

Image resizing: All images are resized to [224,224] using ImageDataGenerator to fit in the model.

*Dataset Labeling*

Each screenshot image was labeled using a software tool named WebScore AI that evaluates the websites. WebScore AI provides the overall evaluation result in the form of scores from 1 to 10. Besides, it shows three different colors with the score which are red, yellow, and green. Red color for low scores, yellow for intermediate scores, and green for high scores. This tool evaluates the websites on a scale from 1 to 10 whereas this project evaluates on a scale from very bad to excellent. As a result, the range from 1 to 10 was divided to be compatible with the labels of this project. In the tool, the range of score from 1 to 3.59 is for red color and it was excluded as a result of several evaluated websites that took a score in that range were not reachable due to denied access and security protection. The range of 3.60 – 5.59 scores that is for yellow color is divided to cover very bad and bad grades. Therefore, the range of 3.60 – 4.59 become indicating very bad, 4.60 – 5.59 indicating bad. Moreover, the range of 5.60 – 10 scores that is for green color is divided to cover good, very good, and excellent. Then, the range 5.60 – 7 indicating good, 7.01 – 8.59 indicating very good, 8.60 – 10 indicating excellent.

*Dataset Splitting*

The dataset is split manually, 70% of images (217 websites) in the train folder, and 30% of images (93 websites) in the test folder. Furthermore, the dataset is imbalanced. So, from each class, 30% of websites are taken for testing and the rest 70% of websites for training.

*Models Implementation*

CNN model has been trained using three convolutional layers with a ReLU activation function. Each of them is followed by a MaxPooling layer. Dropout is used after them. Finally, two fully connected layers with ReLU and SoftMax activations respectively. The input size of the model is 200*200 RGB images. The parameters were used in the model are Adam as an optimizer, Focal loss as a loss, batch size is 32, and epochs equal to 10.

*Models Performance and Discussion*

In this model, the Confusion matrix is used as in the SVM model to measure the model performance. The CNN model achieved an overall accuracy of 69%. Table 2 summarizes the additional performance metrics, including average Precision, Recall, and F1-Score.

Table 2: CNN Model Performance Metrics

| Accuracy | Precision | Recall | F1-Score |
|---|---|---|---|
| 69% | 44% | 34% | 37% |

The Precision score of 44% indicates that the model correctly identified a little under half of the true positive instances. The Recall score of 34% suggests that the model was able to capture only about one-third of the actual positive cases. The F1-Score, which balances Precision and Recall, reflects a moderate overall performance at 37%.

These results suggest that while the CNN model demonstrated reasonable accuracy, there is room for improvement in its ability to consistently and accurately classify website usability based on the provided screenshot dataset. Further refinement of the model architecture, hyperparameters, and/or the training dataset may be necessary to enhance the model's predictive capabilities for this task.

## 7. Conclusion

This study presented an evaluation of website usability through the development and assessment of two intelligent models: SVM model and CNN model. The SVM model evaluated website usability based on four key principles: load time, mobile-UI, high-resolution photos, and contact information. This model demonstrated excellent performance, achieving an outstanding accuracy of 99%. In contrast, the CNN model, which relied on visual analysis of website screenshots, yielded a lower performance, achieving a maximum accuracy of 69%.

The superior performance of the SVM model underscores the effectiveness of leveraging specific, interpretable usability features to assess website quality. The relatively lower accuracy of the CNN model suggests that while visual design elements are important, a more holistic evaluation incorporating both textual and visual features may be necessary to fully capture website usability.

Overall, the findings of this study highlight the potential of advanced machine learning techniques to provide reliable and accurate assessments of website usability. These insights can inform the development of more user-centric online experiences, ultimately enhancing the quality and effectiveness of e-commerce platforms.

## 8. Future Work

The following works are set aside as future work for improving the two models' performance.

- Develop a hybrid model that integrates visual features extracted through deep learning techniques with traditional machine learning algorithms for comprehensive evaluation.
- Explore the integration of multimodal data sources to further enhance the assessment of website usability.
- Expand the dataset size and incorporate additional website features for evaluation.

**Author Contributions:** Conceptualisation, A.H, R.A, A.M and N.A.; methodology, A.H, R.A, A.M and N.A.; software, A.H, R.A and A.M; validation A.H, R.A, A.M and N.A.; formal analysis, A.H, R.A, A.M and N.A.; investigation, A.H, R.A, A.M and N.A.; resources, A.H, R.A and A.M; data curation, A.H, R.A and A.M; writing—original draft preparation, A.H, R.A, A.M and N.A.; writing—review and editing, A.H, R.A, A.M and N.A.; visualisation, A.H, R.A, A.M and N.A. All authors have read and agreed to the published version of the manuscript.

**Funding:** The author received no financial support for this study or the publication of this manuscript.

**Institutional Review Board Statement:** Not applicable.

**Informed Consent Statement:**  Not applicable.

**Data Availability Statement:**  Data is available upon reasonable request from the corresponding author.

**Conflicts of Interest:** The author declares no conflicts of interest.

## References

[1] Bevan, N., Carter, J., Earthy, J., Geis, T., and Harker, S. (2016). New iso standards for usability, usability reports and usability measures. In International conference on human-computer interaction, pages 268–278. Springer.

[2] Ilbahar, E. and Cebi, S. (2017). Classification of design parameters for ecommerce websites: A novel fuzzy kano approach. Telematics and Informatics, 34(8):1814–1825.

[3] Nizamani, S., Khoumbati, K., NIZAMANI, S., Memon, S., and NIZAMANI, S. (2019). Usability evaluation of the top 10 universities of pakistan through guideline scoring. Sindh University Research Journal-SURJ (Science Series), 51(01):151–158.

[4] Dingli, A. and Cassar, S. (2014). An intelligent framework for website usability. Advances in Human-Computer Interaction, 2014.

[5] Jayanthi, B. and Krishnakumari, P. (2016). An intelligent method to assess webpage quality using extreme learning machine. International Journal of Computer Science and Network Security (IJCSNS), 16(9):81.

[6] Khani, M. G., Mazinani, M. R., Fayyaz, M., and Hoseini, M. (2016). A novel approach for website aesthetic evaluation based on convolutional neural networks. In 2016 Second International Conference on Web Research (ICWR), pages 48–53. IEEE.

[7] Lu, H., Wang, L., Ye, M., Yan, K., and Jin, Q. (2018). Dnnbased image classification for software gui testing. In 2018 IEEE SmartWorld, Ubiquitous Intelligence & Computing, Advanced & Trusted Computing, Scalable Computing & Communications, Cloud & Big Data Computing, Internet of People and Smart City Innovation (SmartWorld/SCALCOM/UIC/ATC/CBDCom/IOP/SCI), pages 1818–1823. IEEE.

[8] Jafary, H., Taghavifard, M. T., Hanafizadeh, P., and Kazazi, A. (2018). Intelligent quality assessment model of news sites (newsqual). International Journal of Web Research, 1(1):51–61.

[9] Dou, Q., Zheng, X. S., Sun, T., and Heng, P.-A. (2019). Webthetics: quantifying webpage aesthetics with deep learning. International Journal of HumanComputer Studies, 124:56–66.

[10] Ghattas, M. M. and Sartawi, P. D. B. (2020). Performance evaluation of websites using machine learning.

[11] Singh, T., Malik, S., and Sarkar, D. (2016). E-commerce website quality assessment based on usability. In 2016 International Conference on Computing, Communication and Automation (ICCCA), pages 101–105. IEEE.

[12] Hustak, T. and Krejcar, O. (2016). Principles of usability in human-computer interaction. In Advanced Multimedia and Ubiquitous Engineering, pages 51–57. Springer.

[13] Majid, E. S. A., Kamaruddin, N., and Mansor, Z. (2015). Adaptation of usability principles in responsive web design technique for e-commerce development. In 2015 International Conference on Electrical Engineering and Informatics (ICEEI), pages 726–729. IEEE.

[14] Mvungi, J. and Tossy, T. (2015). Usability evaluation methods and principles for the web. International Journal of Computer Science and Information Security, 13(7):86.

[15] Nielsen, J. (1994). Enhancing the explanatory power of usability heuristics. In 87 Proceedings of the SIGCHI conference on Human Factors in Computing Systems, pages 152–158.

[16] Joshi, R. (2016). Accuracy, precision, recall f1 score: Interpretation of performance measures. Available at https://blog.exsilio.com/tag/ accuracy/. Accessed: 2021-3-7.